\documentclass[aps,pra,twocolumn,showpacs,groupedaddress]{revtex4-1}
\usepackage{graphicx}  
\usepackage{dcolumn}   
\usepackage{bm}        
\usepackage{amssymb}   
\usepackage{amsmath}
\usepackage{braket}
\usepackage{enumitem}
\usepackage[mathscr]{euscript}
\usepackage{color}
\usepackage{epsfig,color}
\usepackage{hyperref}
\DeclareUnicodeCharacter{3B1}{\ensuremath{\alpha}}

\begin{document}

\title{ICO learning as a measure of transient chaos in  $\mathcal{PT}$-symmetric Li\'enard systems}

\author{J. P. Deka}
\email{jyotiprasad$_$physics@gcuniversity.ac.in}
\affiliation{Department of Physics, Girijananda Chowdhury University, Guwahati - 781017, India}

\author{A. Govindarajan}
\email{govin.nld@gmail.com}
\affiliation{Department of Nonlinear Dynamics, School of Physics, Bharathidasan University, Tiruchirappalli - 620024, India}

\author{A. K. Sarma}
\email{aksarma@iitg.ac.in}
\affiliation{Department of Physics, Indian Institute of Technology Guwahati, Guwahati - 781039, India}

\date{\today}

\begin{abstract}
	In this article, we investigate the implications of the unsupervised learning rule known as Input-Correlations (ICO) learning in the nonlinear dynamics of two linearly coupled $\mathcal{PT}$-symmetric Li\'enard oscillators. The fixed points of the oscillator have been evaluated analytically and  the Jacobian linearization is employed to study their stability. We find that on increasing the amplitude of the external periodic drive, the system exhibits period-doubling cascade to chaos within a specific parametric regime wherein we observe emergent chaotic dynamics. We further notice that the system indicates an intermittency route to chaos in the chaotic regime. Finally, in the period-4 regime of our bifurcation analysis, we predict the emergence of \textit{transient chaos} which eventually settles down to a period-2 oscillator response which has been further validated by both the maximal Finite-Time Lyapunov Exponent ($\mathcal{FTLE}$) using the well-known Gram-Schmidt orthogonalization technique and the Hilbert Transform of the time-series. In the transiently chaotic regime, we deploy the ICO learning to analyze the time-series from which we identify that when the chaotic evolution transforms into periodic dynamics, the synaptic weight associated with the time-series of the loss oscillator exhibits stationary temporal evolution. This signifies that in the periodic regime, there is no overlap between the filtered signals obtained from the time-series of the coupled $\mathcal{PT}$-symmetric oscillators. In addition, the temporal evolution of the weight associated with the stimulus mimics the behaviour of the Hilbert transform of the time-series.
\end{abstract}

\maketitle

\section{Introduction}

In neuroscience, Hebbian learning is an attempt to explain the synaptic efficacy when the post-synaptic cells are continuously stimulated by the pre-synaptic cells [1]. In this context, it could be quoted that `\textit{Neurons that fire together wire together}' [4].  Synaptic plasticity plays an important role in the coordinated implementation of cells that fire in synchrony. Furthermore, there is an evidence that synaptic strength could either be strengthened or weakened due to heterosynaptic plasticity. In this context, the \textit{ICO learning rule} is a method that incorporates the rules of \textit{classical Hebbian learning} (CHL) along with the conditions of \textit{heterosynaptic plasticity} [1-3,5-6, 44]. It takes into account the correlation between the synapses and based on this, the \textit{synaptic strength} (measured in terms of the weight) could be strengthened or weakened. In addition, this rule is a variant of \textit{differential Hebbian learning} and it takes into account the changes in the post-synaptic cells. One major advantage of this method is that it circumvents the potentially destabilising term in CHL by replacing the derivative of the output of the learning rule with the derivative of the input. This enables the use of higher learning rates together with the removal of the destabilising term in the learning rule.

On the other hand, Carl M. Bender and his graduate student Stefan Boettcher's ground-breaking discovery [7] that non-Hermitian Hamiltonians can yield a real eigenspectra paved the path for a new era in the foundational studies of quantum mechanics. Such Hamiltonians were termed as $\mathcal{PT}$-symmetric Hamiltonians and as such, they are invariant under the joint operation of parity ($\mathcal{P}$) and time-reversal ($\mathcal{T}$) operators by commuting with the $\mathcal{PT}$-operator $[H,PT]=0$. Unlike Hermitian Hamiltonians, such non-Hermitian Hamiltonians possess an \\exceptional-point ($\mathcal{EP}$) which is accompanied by the so-called $\mathcal{PT}$-symmetric phase transition in the eigenspectra after which the former changes from real to imaginary. The regime where the eigenspectrum is real is known as  unbroken $\mathcal{PT}$-symmetric regime and if the eigenspectrum is purely imaginary, then the regime is called as the broken $\mathcal{PT}$-symmetry.

In the context of experimental endeavors, El-Ganainy \textit{et al.} [8] were the first to put forward the proposition that evanescently coupled dielectric optical slab-based waveguides with balanced gain and loss can facilitate a feasible way to  validate these mathematical theories in optics. Such a proposition was deemed conceivable due to the fact that the time-independent Schr\"odinger equation and the paraxial equation of diffraction/dispersion are isomorphic. The potential function in the former and the refractive index in the latter serve as the connecting tool for this shared isomorphic trait. Not long ago, the first experimental observation was reported in a coupled optical waveguide configuration by R\"uter \textit{et al.} [8] followed by Guo \textit{et al.} [9]. Two optical waveguides were etched on to a lithium niobate  substrate. One of the channels was optically pumped to provide amplification while the other waveguide was provided an equal proportion of attenuation. In such a $\mathcal{PT}$-symmetric optical waveguide configuration, the $\mathcal{PT}$ phase transition was characterized by a transition from periodically evolving optical power to exponential growth and decay of optical power in the two channels. In the context of optical systems, the notion of $\mathcal{PT}$-symmetry has been explored in diverse domains such as optoelectronic oscillators [10], optomechanical systems [11-13], modulational instability in complex media [14], active LRC circuits [15], multilayered structures [16-17], unidirectional invisibility in periodic structures [18-19], wireless power transfer [20], optical lattices [21-24], double ring resonators [25], and so on.

The temporal dynamics of most continuous nonlinear systems are usually either asymptotically stable (otherwise oscillatory) or unstable. Oscillatory temporal dynamics, on the other hand, could either be periodic, quasiperiodic or chaotic. But in transient chaotic dynamics, the temporal dynamics undergo a transition in their behavior from chaotic to periodic as the system is allowed to evolve. Hence, it can be inferred that chaotic dynamics that arise for a finite length of time can be termed as transient chaos [26-31]. Grebogi, Ott and Yorke's discovery of the emergence of transient chaotic dynamics in certain systems passing through crisis led to the birth of this field in nonlinear dynamics [29]. Moreover, Cruthfield and Kaneko's work on the interconnection between attractors and turbulence (spatiotemporal complexity) generated a new perspective on the study of turbulence. Such phenomena are related to attractors or long-lived transient dynamics of non-attracting chaotic sets [30]. Subsequently, transient chaos has been reported in electronic oscillators [32-34], lasers [35-37], parametrically forced pendulum [38], Duffing oscillators [39] and so on. Quantification of such chaotic dynamics requires the computation of the maximal $\mathcal{FTLE}$, which has been observed to decrease from a positive value to zero, thereby signifying the emergence of a limit cycle attractor [39].

In this article, we report the transient chaotic dynamics that emerge in  a $\mathcal{PT}$-symmetric coupled Li\'enard systems for non-autonomous excitation. The article is organized as follows. In section II, we present a brief overview of the mathematical model. This is followed by an analysis of the stability of the fixed points using the Jacobian linearization and how transient chaotic dynamics arise in our system, which we have further validated by studying the maximal $\mathcal{FTLE}$ and the Hilbert transform of the time-series in section III. And finally, in Section IV, we investigate the implication of the unsupervised learning rule known as ICO learning in the analysis of time-series transformation from chaotic to peridic evolution, followed by our conclusion in Sec. V.

\section{\label{sec2}MATHEMATICAL MODEL}
Let us consider the $\mathcal{PT}$-symmetric Li\'enard oscillator given below.
\begin{equation}
	\ddot{x}+f(x) \dot{x}+\alpha x+g(x)=0
\end{equation}
where $\alpha$ is a real constant, $f(x)$ is the dissipation/amplification profile and $g(x)$ is the nonlinearity profile of the oscillator. Under the joint operation of the $\mathcal{PT}$-operator ($x\rightarrow-x$ and $t\rightarrow-t$), one can show that the above equation gets transformed as follows
\begin{equation}
	\ddot{x}-f(-x)\dot{x}+\alpha x-g(-x)=0
\end{equation}
Now, if $f(x)$ and $g(x)$ are both anti-symmetric in $x$, then the oscillator whose dynamics is governed by Eq. (1) is $\mathcal{PT}$-symmetric. So, in this article, we consider two configurations of coupled $\mathcal{PT}$-symmetric oscillators with dissipation profile $f(x_i)=\eta x_i$ and nonlinearity profile $g(x_i)=\beta x_i^3$ (where $i=1,2$) given below
\begin{subequations}
	\begin{align}
		& \ddot{x}_1+\eta x_1 \dot{x}_1+\alpha x_1+\beta x_1^3+\kappa x_2=0 \\
		& \ddot{x}_2+\eta x_2 \dot{x}_2+\alpha x_2+\beta x_2^3+\kappa x_1=0
	\end{align}
\end{subequations}
Here, $\kappa$ is the coupling constant. Also, the term $\eta$ denotes the strength of nonlinear dissipation/amplification and $\beta$ refers to the coefficient of Duffing nonlinearity of the two oscillators. It may be noted here that the dissipation/amplification profile $f(x_i)=\eta x_i$ could provide both dissipation and amplification in the same oscillator depending on whether the parameter $\eta$ is positive or negative. It must also be noted  that in this autonomous case, the individual oscillators are $\mathcal{PT}$ symmetric, but the coupled oscillator configuration, as a whole, is not $\mathcal{PT}$ symmetric.

In the context of an external agent driving the system, the mathematical model will be transformed as follows.
\begin{subequations}
	\begin{align}
		& \ddot{x}_1 - \eta x_1 \dot{x}_1 + \alpha x_1 + \beta x_1^3 + \kappa x_2 = A_0 cos(\omega t)\\
		& \ddot{x}_2 + \eta x_2 \dot{x}_2 + \alpha x_2 + \beta x_2^3 + \kappa x_1 = A_0 cos(\omega t)
	\end{align}
\end{subequations}
where $A_0$ is the amplitude of the external drive and $\omega$ is the frequency of the drive. One can easily observe that this system remains invariant under the joint operation of the parity operator defined as $\mathcal{P}: x_1 \leftrightarrow x_2$ and the time-reversal operator defined as $\mathcal{T}: t \rightarrow -t$ and hence, it is obvious that this coupled oscillator configuration is $\mathcal{PT}$-symmetric. It must also be noted that the choice of the external drive is vital in such a scenario and as such, we have chosen a cosine function drive which is an even function of time. Now, since the amplification/dissipation profile chosen in both configurations is similar, in the second configuration, the term signifying gain/loss in one of the oscillators should carry the opposite sign so as to play the antithetical role. Hence, we will term the first system as the gain and the second as the loss oscillator in our results and discussion section. In our analysis, to excite the gain oscillator we have chosen the initial condition as, $(x_1, y_1, x_2, y_2) = (0.1, 0, 0, 0)$, where $y_i = \dot{x}_i$ and $i = 1, 2$.

\section{\label{sec3}ANALYSIS OF FIXED POINTS AND TIME-SERIES}
Before analyzing the coupled oscillator configuration, we would like to study the stability of the fixed points of the first oscillator in the absence of coupling for the autonomous case. By setting $\dot{x}_1=y_{1}$ and $\kappa=0$, we have
\begin{subequations}
	\begin{align}
		& \dot{x}_1= y_1 \\
		& \dot{y}_1= -\eta x_1 y_1- \alpha x_1- \beta x_1^3
	\end{align}
\end{subequations}
Hence the fixed points (FP) of this nonlinear system are read as
\begin{enumerate}
	\item FP1 $\rightarrow$ $(x_1,y_{1})=(0,0)$
	\item FP2 $\rightarrow$ $(x_1,y_{1})=(\sqrt{-\alpha / \beta},0)$
	\item FP3 $\rightarrow$ $(x_1,y_{1})=(-\sqrt{-\alpha / \beta},0)$,
\end{enumerate}
\vspace{1.0em}
\begin{figure}
	\centering
	\includegraphics[width=0.8\linewidth]{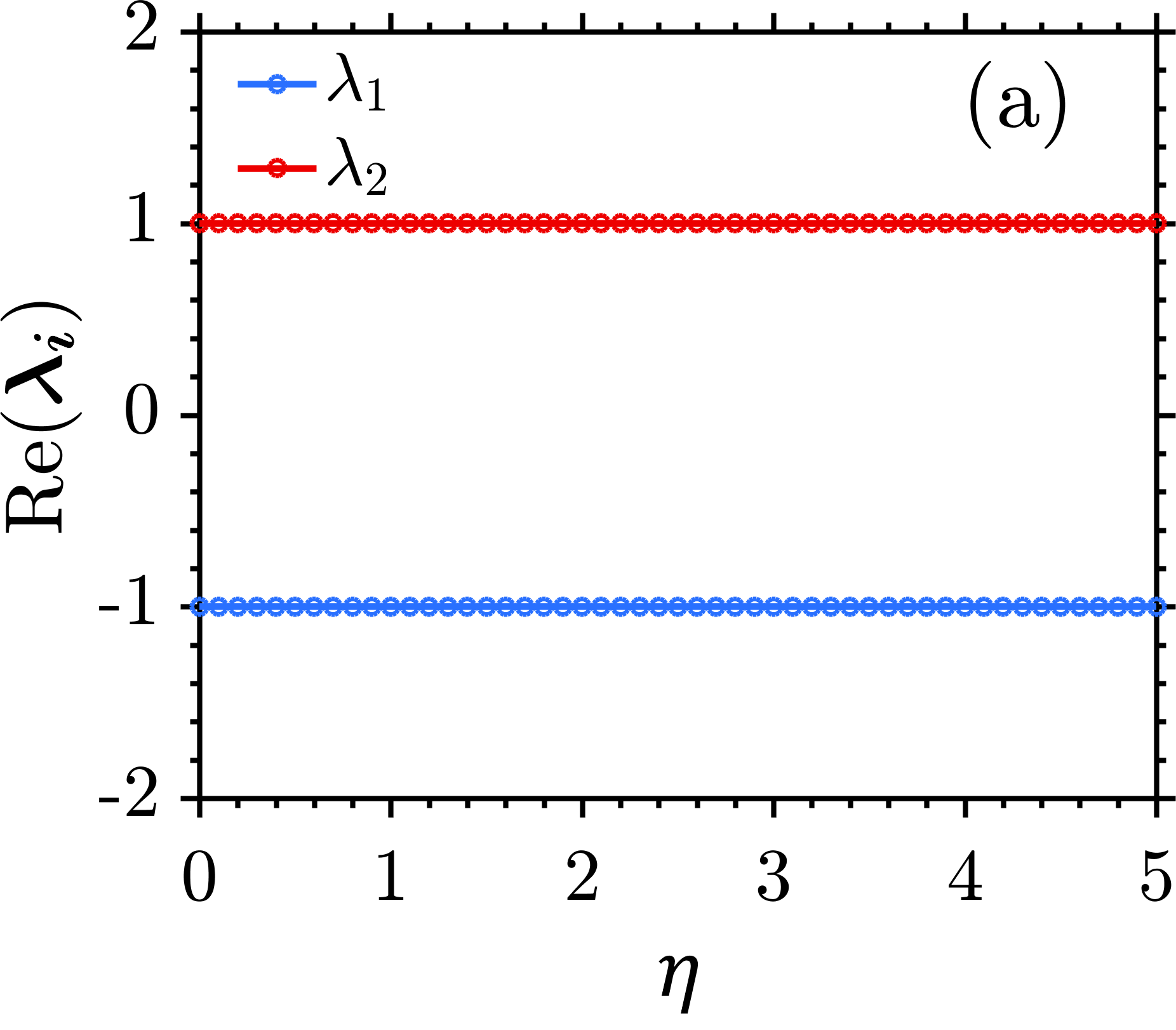}
	\includegraphics[width=0.8\linewidth]{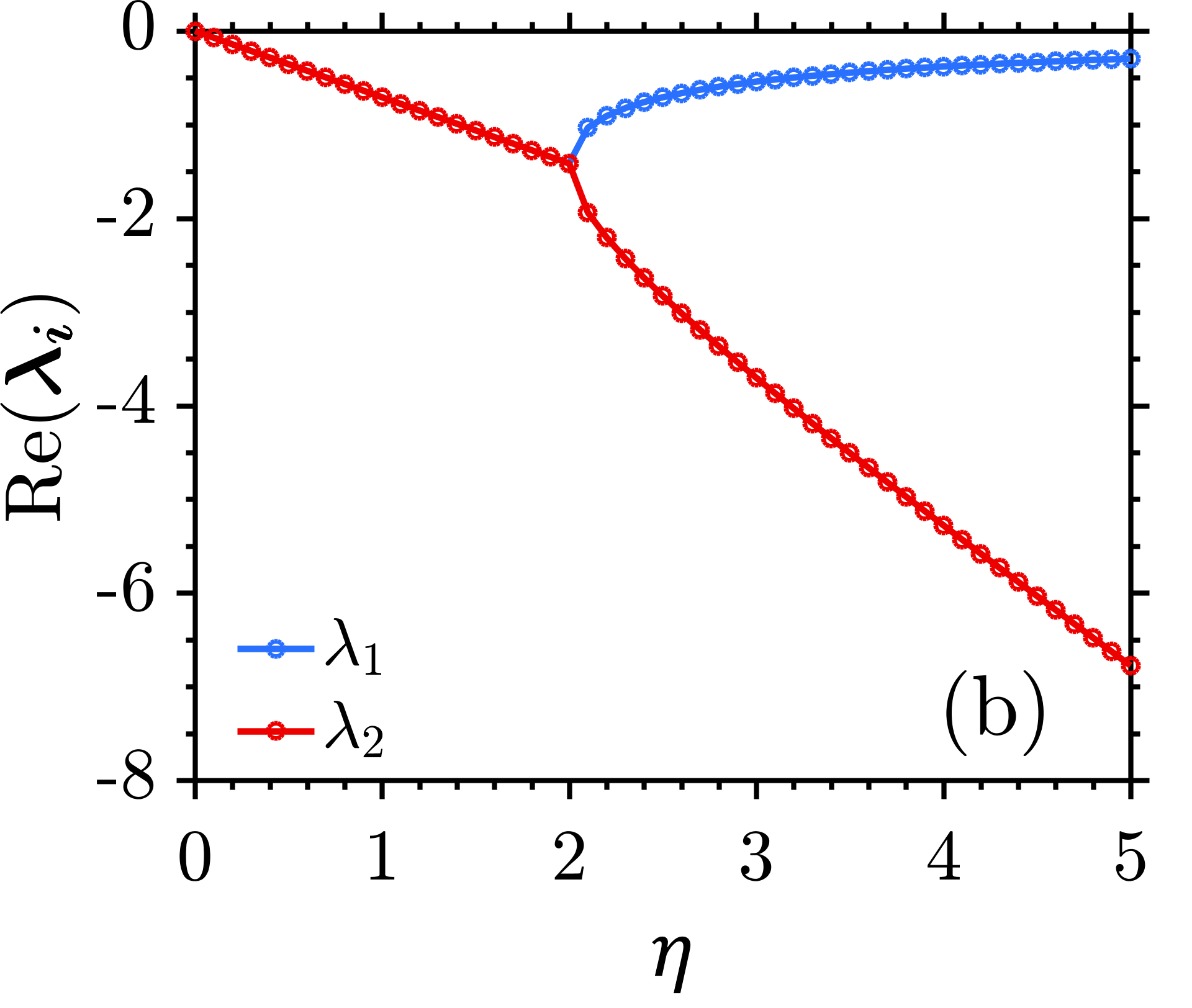}
	\includegraphics[width=0.8\linewidth]{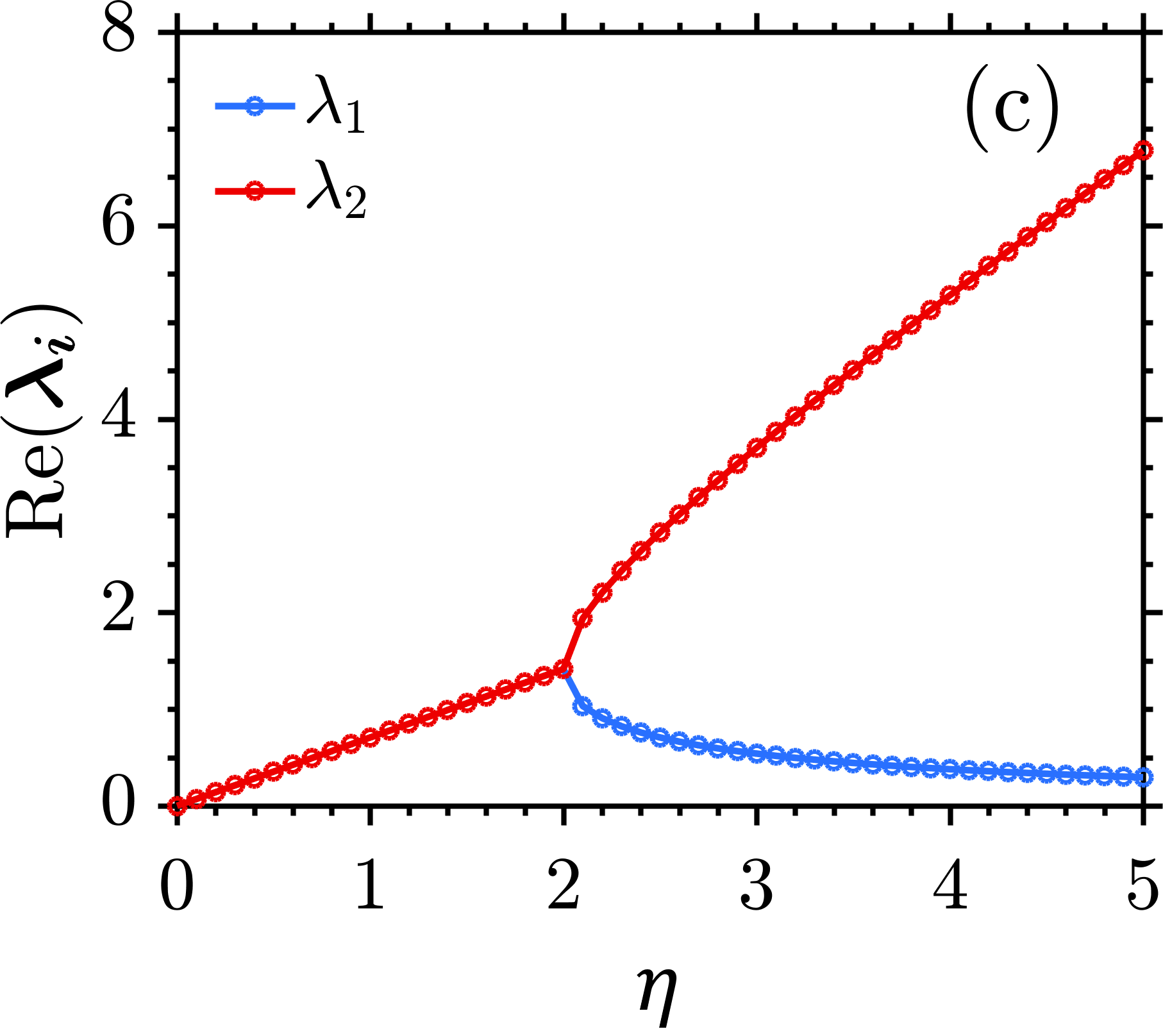}
	\caption{Eigenspectra of the  Jacobian linearization of (a) FP1 (b) FP2 and (c) FP3 for $\beta=0.5$ and $\alpha=-1$}
\end{figure}
Spectral analysis of the Jacobian eigenspectra shows that FP1 is a saddle fixed point, FP2 is a stable fixed point and FP3 is an unstable fixed point, which have been depicted clearly in Fig. 1. One interesting feature that one can observe from our spectral analysis is that our system contains all three forms of hyperbolic fixed points. Moreover, it is to be noted that the potential function for a single oscillator is given by $V(x)=\alpha x_{1}^2/2 + \beta x_{1}^4/4$ and stability analysis of the equilibrium points of this potential attests that the equilibrium point $x = \pm \sqrt{-\alpha/\beta}$ is stable if $\alpha<0$ and $\beta>0$. Generally, we have a double well potential for certain choice of parameters. But the presence of nonlinear position dependent dissipation renders one of the stable equilibrium points unstable. Furthermore, another interesting aspect to be noted here is that the eigenspectra of all three fixed points exhibit no exceptional point (EP). Normally, in a $\mathcal{PT}$-symmetric system, the eigenspectra undergo a phase transition from real to imaginary eigenvalues as certain parameters (like gain/loss) are varied. But in this configuration where the position dependent dissipation ($\eta$) makes the system to be $\mathcal{PT}$-symmetric, there is no such a phase transition in the eigenspectra of the Jacobian linearization  albeit the system obeys the so-called $\mathcal{PT}$-symmetry. A complete and detailed analysis of the eigenspectra for the linearization Jacobian for this system has been presented in our previous work [43].

We would now like to start our discussion with the bifurcation diagram of the temporal maxima of the system under study versus the amplitude of the external drive $A_0$. It can be clearly seen in Fig. 2 that on increasing the value of $A_0$, the temporal dynamics of the gain oscillator transforms from periodic to chaotic. We have considered a very lengthy time-series for our bifurcation analysis, ranging from $t=0$ to $t=25000$, out of which we have removed the temporal data up to $t=23000$ as transient. We will now concentrate on a few specific cases of the amplitude of the external drive. Some of the regimes in the gain oscillator response to the amplitude of the external drive are shown in Table I. The phase plane of the oscillator response for this parametric regime is shown in Fig. 3. In our analysis, we have chosen the following parameter values: $\eta=0.1$, $\beta=0.5$, $\kappa=0.5$, $\alpha=0.25$ and $\omega=0.25$. The amplitude of the external drive $A_0$ is chosen as the variable parameter.
\begin{figure}
	\centering
	\includegraphics[height=6cm,width=7cm]{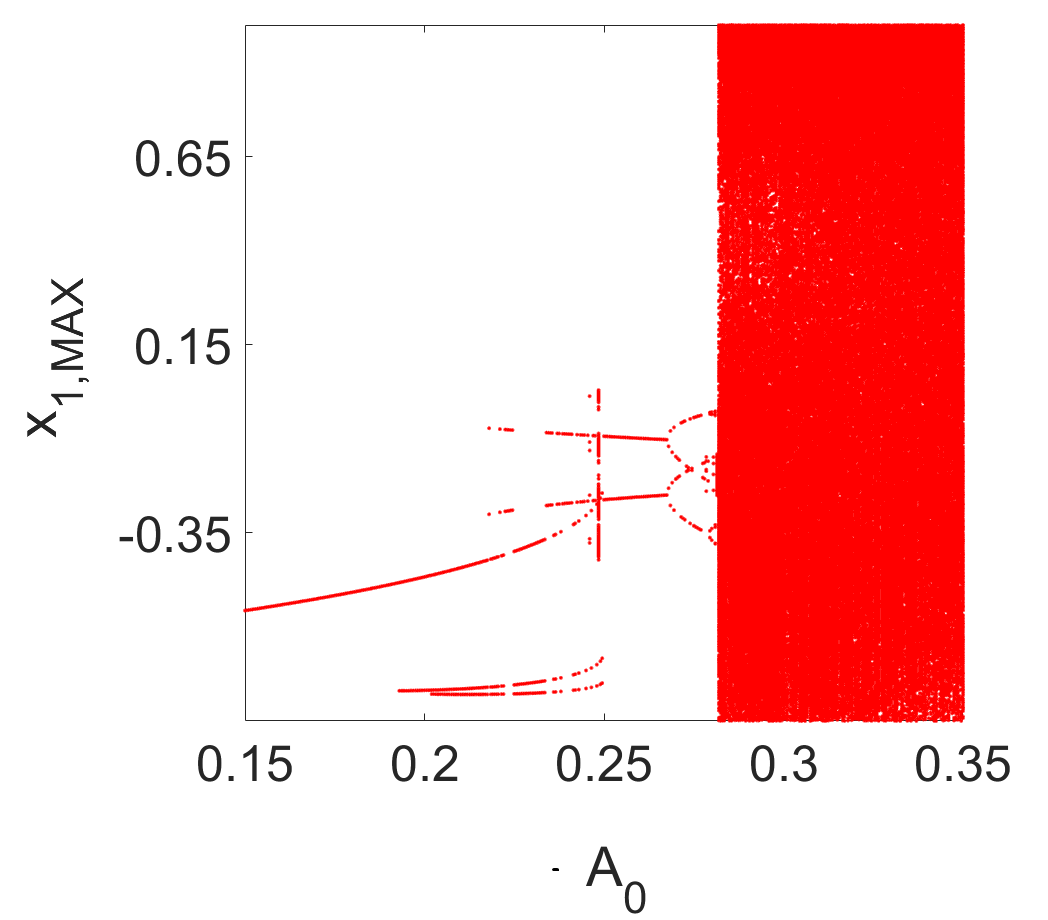}
	\caption{Temporal maxima of the gain-oscillator $x_{1,max}$ vs. $A_0$.}
\end{figure}
\begin{table}
	\centering
	\includegraphics[height=3cm,width=7cm]{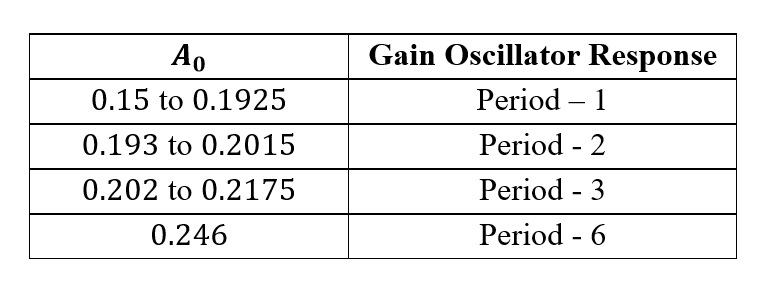}
	\caption{Temporal response of the gain oscillator.}
\end{table}
It can be clearly seen from Fig. 4 that as we increase the value of $A_0$, the gain oscillator starts to exhibit multiple temporal maxima and beyond a certain threshold, we observe the emergence of chaotic dynamics in the oscillators. This has also been portrayed through the phase plane analysis in Fig. 3. We would now like to present an in-depth analysis of the parametric regime before the system transits into chaotic dynamics. One can notice from Fig. 4 that the system exhibits a case of period doubling cascade to chaos. But at the same time, the bifurcation analysis also depicts chaotic dynamics for $A_0=0.27148$ and $A_0=0.27772$ and a period-2 and a period-6 oscillator response for $A_0=0.2784$ and $A_0=0.27852$, respectively. This means even within the periodic regime, there are regions where the oscillators might portray chaotic or period-N temporal dynamics ($N \geq 2$). This has been depicted in the time-series of the gain oscillator shown in Fig. 5. Hence, we can draw a conclusion that our system exhibits intermittency in the bifurcation analysis depicting the period doubling cascade to chaos.

\begin{figure}
	\centering
	\includegraphics[height=7cm,width=8cm]{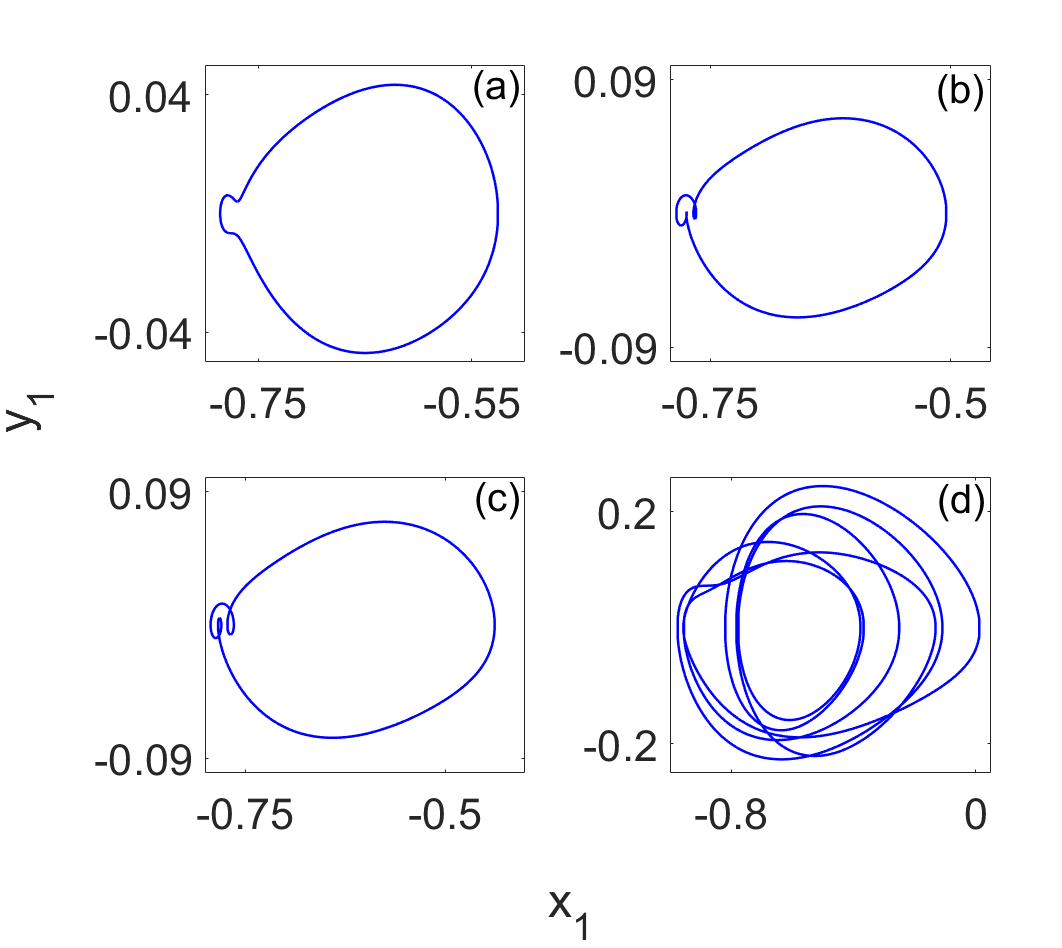}
	\caption{Phase plane of the gain-oscillator for (a) $A_0=0.17$, (b) $A_0=0.205$, (c) $A_0=0.212$ and (d) $A_0=0.246$.}
\end{figure}

\begin{figure}
	\centering
	\includegraphics[height=6cm,width=7cm]{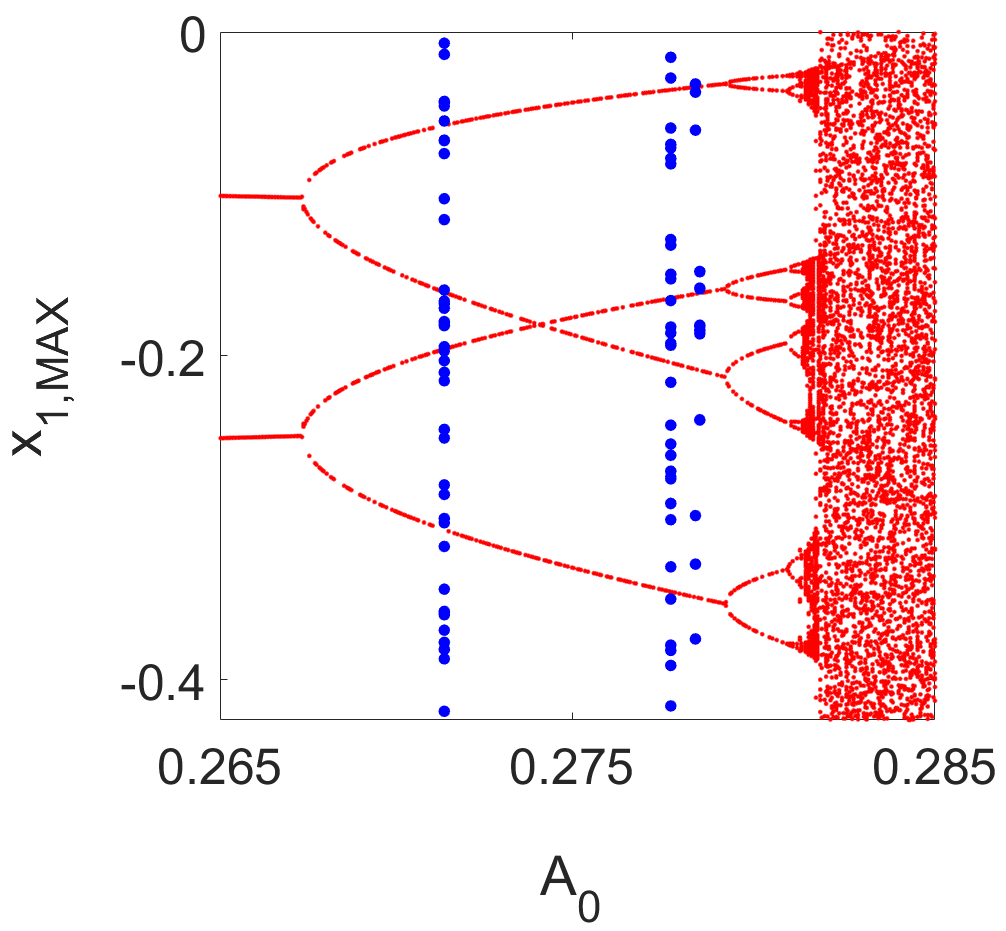}
	\caption{Temporal maxima of the gain-oscillator delineating the period doubling cascade to chaos.}
\end{figure}

\begin{figure}
	\centering
	\includegraphics[height=6cm,width=7cm]{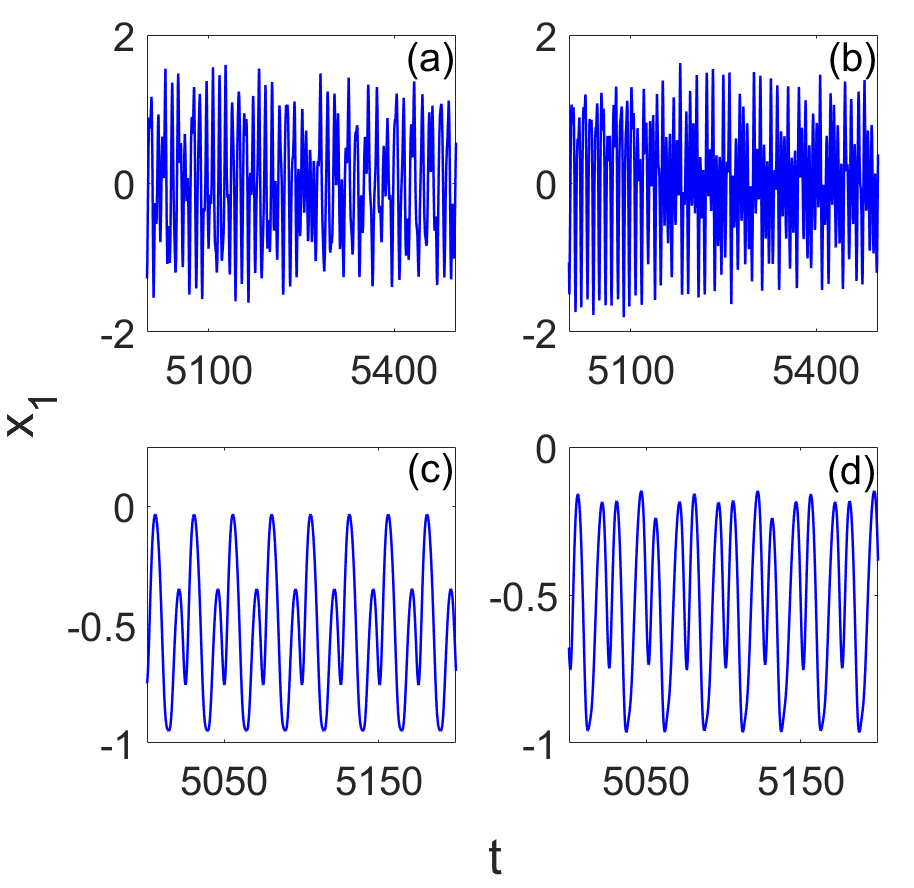}
	\caption{Temporal evolution of the gain-oscillator for (a) $A_0=0.27148$, (b) $A_0=0.27772$, (c) $A_0=0.2784$ and (d) $A_0=0.27852$.}
\end{figure}

Also, from Fig. 6, we can observe that the temporal evolution delineate intermittent chaotic dynamics for $A_0=0.282$. This could be attributed to the reason that $x_1=-0.7$ is a stable fixed point for the gain oscillator and this has been shown in the spectral analysis of the  Jacobian linearization in our previous work [43]. So, from the temporal evolution in Fig. 6, one can notice that for the duration between $t=1240$ to $t=1440$, the gain oscillator  exhibits a period-4 oscillatory dynamics about its stable fixed point. Hence, we can conclude that $\mathcal{PT}$-symmetric Li\'enard systems reveal the intermittency route to chaos. Furthermore, intermittency could be easily characterized by studying the temporal maxima of the time-series. In fact, it is the same algorithm that we used for our bifurcation analysis in Fig. 2 and Fig. 4. This has been plotted in Fig. 7 from which one can notice that there is a tiny window from $t=1240$ to $t=1440$ wherein we observe that there is a periodicity in the temporal evolution of $x_{1,max}$.

\begin{figure}
	\centering
	\includegraphics[height=4.5cm,width=8cm]{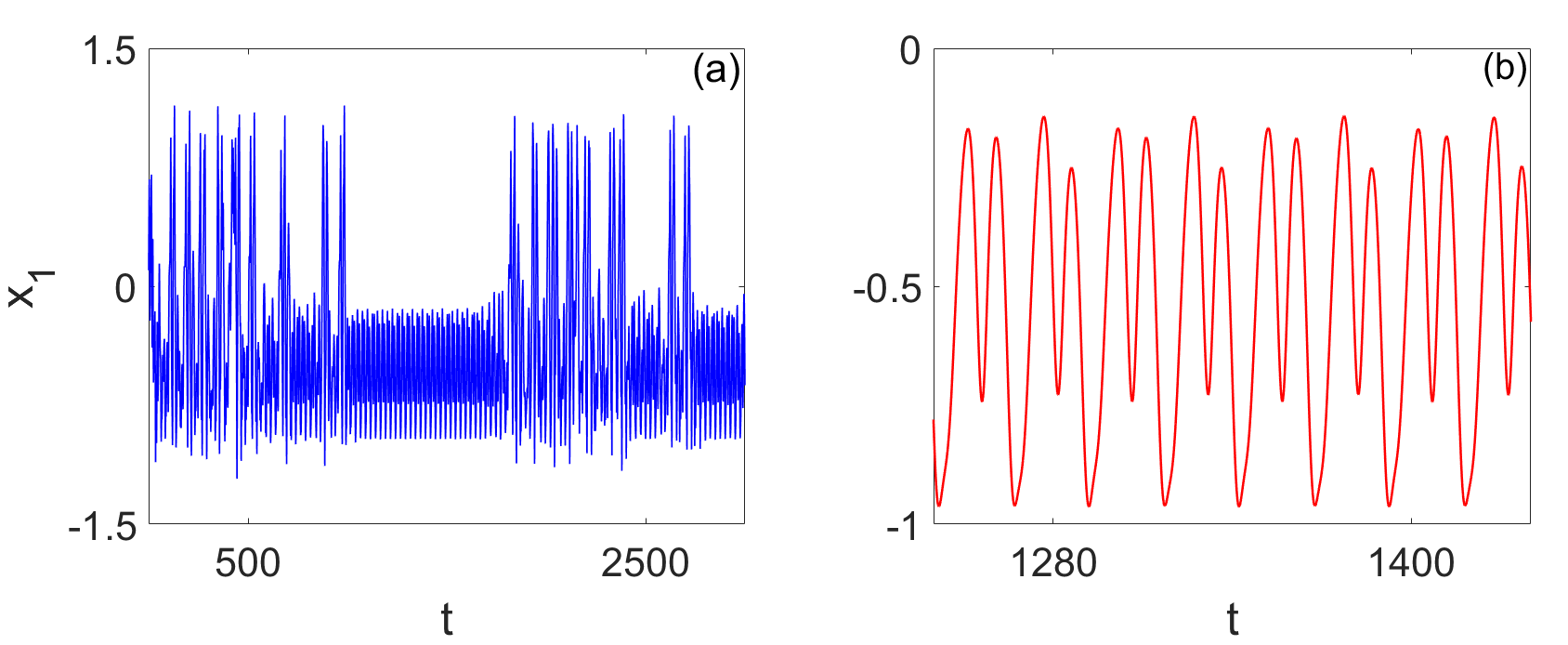}
	\caption{Temporal evolution of the gain-oscillator for $A_0=0.282$.}
\end{figure}

\begin{figure}
	\centering
	\includegraphics[height=6cm,width=7cm]{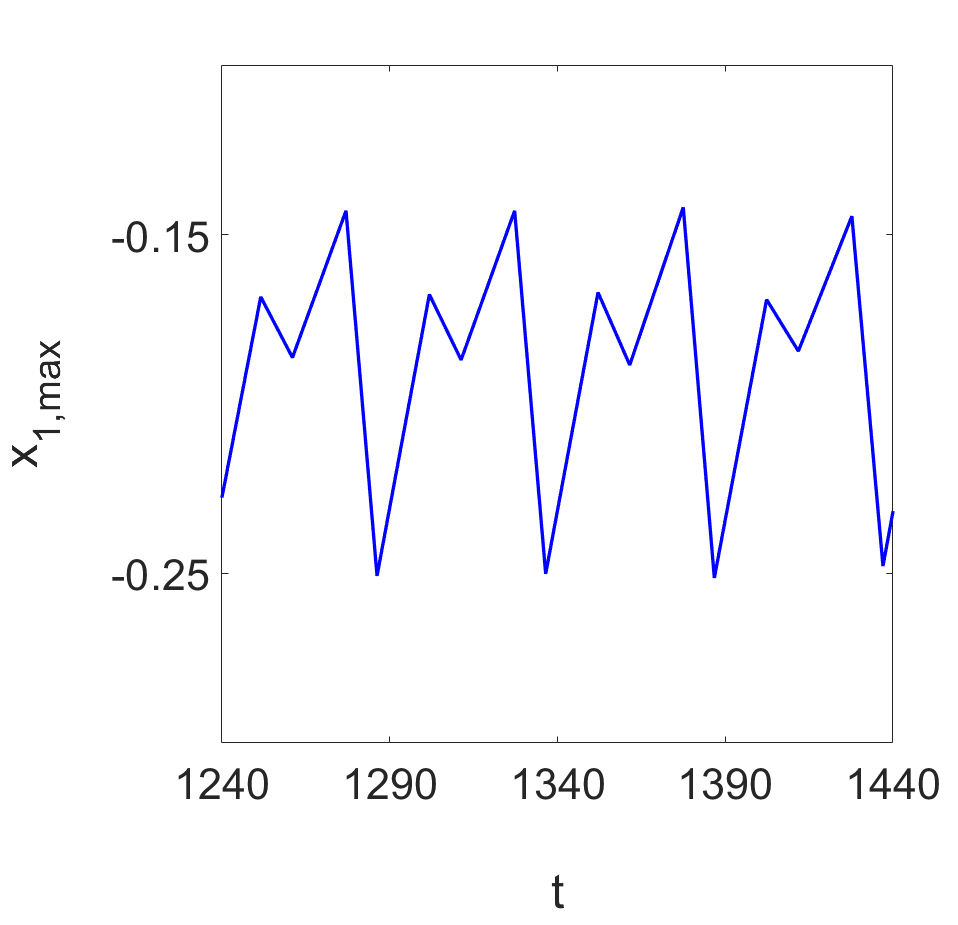}
	\caption{Temporal maxima of the gain-oscillator for $A_0=0.282$.}
\end{figure}

Now, we would like to concentrate on a particular case in the parametric regime of Fig. 4. From Fig. 8(a), one can easily identify that for $A_0=0.274$, the system exhibits chaotic dynamics before it settles down into periodic sinusoidal oscillations. One of the peculiar characteristics that can be perceived from Fig. 8(a) is that the system appears to be in a transiently chaotic state for an incredibly long time. This, in fact, proved quite a tedious computational challenge when the temporal maxima of the time-series were being evaluated for the bifurcation diagrams in Fig. 2 and Fig. 4. This is the reason why we have considered an extremely long time-series for our bifurcation analysis. From Fig. 8(b) and 8(c), it could be gathered that the chaotic phase plane eventually transforms into a period-2 temporal evolution. Transient chaos can be easily characterized by analyzing the maximal $\mathcal{FTLE}$. As the temporal evolution of the two oscillators is allowed to progress, it is observed that the chaotic oscillations eventually cease after a finite length of time and that the time-series cease to be chaotic after $t \approx 5000$ (Fig. 8(a)). This leads us to infer that our system exhibits transient chaotic dynamics. In other words, it means that chaos exists for a finite length of time. Transient chaos can be characterized by studying the maximal $\mathcal{FTLE}$ with a moving window of finite time-series. We have chosen the Gram-Schmidt orthogonalization method to calculate the maximal $\mathcal{FTLE}$ [40-42]. So, accordingly, in our numerical computation, we have chosen to evaluate the $\mathcal{FTLE}$ for a sufficiently long temporal evolution of the system.  Considering these aspects, we have plotted Fig. 8(d) which depicts the maximal $\mathcal{FTLE}$ against the length of the time-series. As observed in the figure,  while we increase the length of $t$, $\mathcal{FTLE}$ exponentially decays and there is an asymptotic convergence of $\mathcal{FTLE}$ towards the $x$-axis. At the same time, if we look at the time-series in Fig. 8(a), one can notice that the chaotic oscillations cease after $t=5000$. Upon further increasing the value of $t$, the asymptotic convergence of $\mathcal{FTLE}$ continues which validates that the oscillations try to converge to a limit cycle attractor. Hence, from our analysis, one can claim that when we have a coupled $\mathcal{PT}$-symmetric Li\'enard system with the same characteristics except for the nonlinear dissipation/amplification and the system is driven by an external agent, the oscillators exhibit the emergence of a chaotic attractor which ceases to exist after a finite length of time as the oscillating temporal dynamics converge to a limit cycle attractor (Fig. 8(c)).
\begin{figure}
	\centering
	\includegraphics[height=8cm,width=9cm]{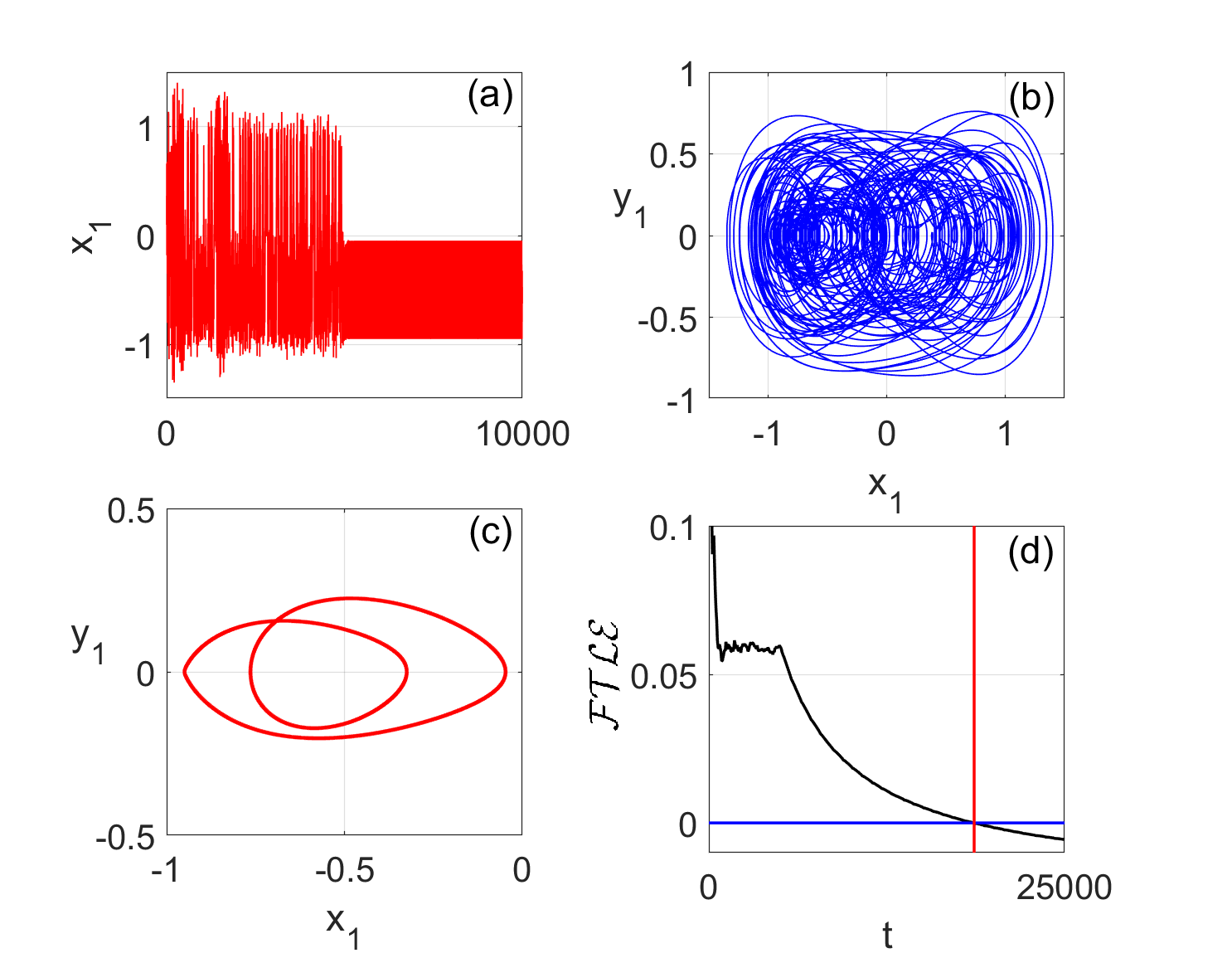}
	\caption{(a) Time-series, (b) transient state phase plane (c) steady-state phase-plane and (d) $\mathcal{FTLE}$ of the gain oscillator for $A_0=0.274$.}
\end{figure}

Transient chaos can further be characterized by studying the Hilbert Transform of the time-series which is given by the following relation,
\begin{equation}
	\hat{H}[x_i(t)]=\frac{1}{\pi}P.V.\int_{-\infty}^{+\infty}\frac{x_i(\tau)}{t-\tau}d\tau 
\end{equation} 
where $x_i(t)$ is the time-series of the system. In Fig. 9, we have plotted the Hilbert Transform of the  time-series shown in Fig. 8(a). From the figure, one can observe that in the regime where the time-series is transiently chaotic, the Hilbert Transform is seen to increase monotonically up to $t\approx5000$ after which the time-series transforms into a periodic temporal evolution. At this junction, the Hilbert Transform yields a constant value as the oscillator exhibits periodic temporal evolution. Thus, we confirm that studying the Hilbert Transform of the time-series serves as a viable way to characterize transient chaos in the nonlinear systems [43].

\begin{figure}
	\centering
	\includegraphics[height=6cm,width=7cm]{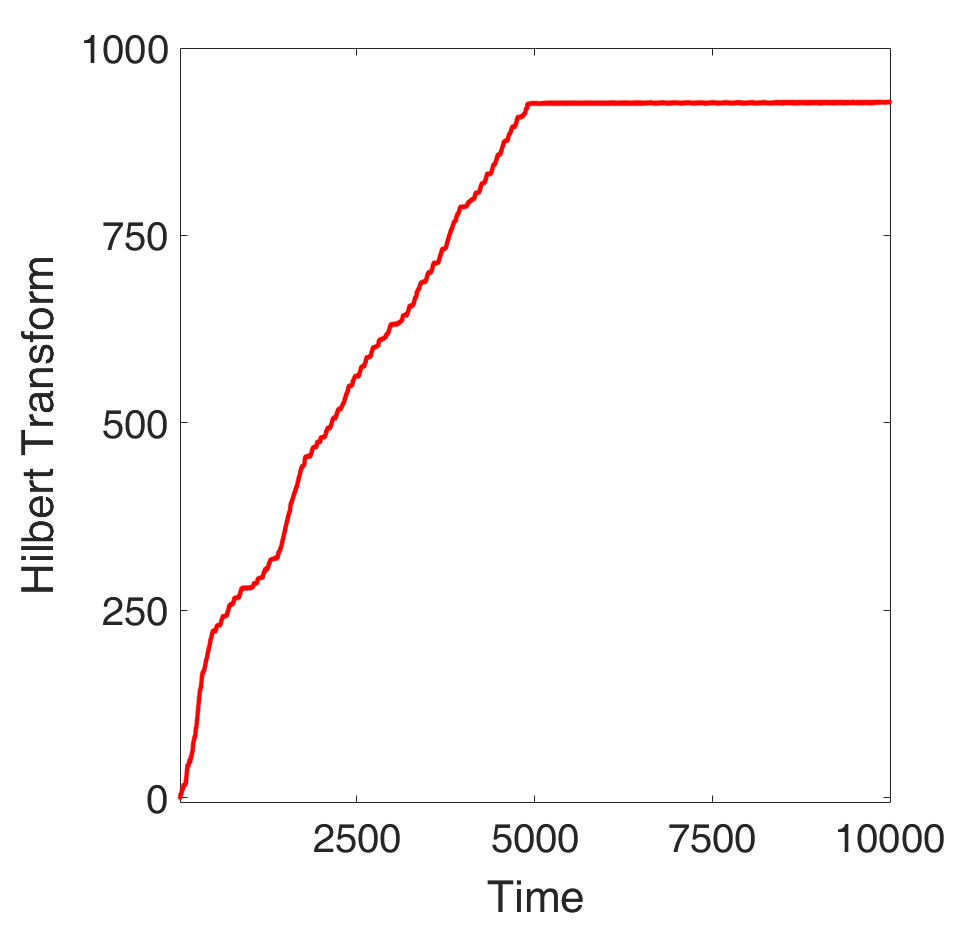}
	\caption{Hilbert transform of the time-series of the gain oscillator for $A_0=0.274$.}
\end{figure}

\section{\label{sec4}ICO Learning}

We shall now explore the implications of the unsupervised learning rule known as \textit{ICO Learning} in our transiently chaotic time-series of the $\mathcal{PT}$-symmetric coupled oscillator system. This rule was first put forward by Prof. Florentin Wörgötter and Prof. Bernd Porr, which is a modification of \textit{Hebbian differential learning} [1-3, 44]. In 2023, the optics group under Prof. Claudio R. Mirasso and Prof. Ingo Fischer of IFISC-UIB, Spain experimentally demonstrated this technique using a pulsed-optoelectronic oscillator [45]. In this rule, the synaptic weights account for the instantaneous change in presynaptic connection by the cross-correlation between the presynaptic events instead of the presynaptic and postsynaptic events. Here, there are two types of input signals - the single reference signal $u_1$ and the multiple stimulus signals $u_{i>1}$ and their corresponding weights are updated by the rule given below.

\begin{subequations}
	\begin{align}
		& \frac{d w_i}{dt} = \eta u_i \frac{du_1}{dt}
	\end{align}
\end{subequations}

where $\eta=0.01$ is the learning rate in our simulations. The weight associated with the reference signal is fixed at $w_1=1$ and the weight for the stimulus $w_{i>1}$ evolves with time and $u_{1}$ and $u_{i>1}$ are the temporally extended spiking signals. This rule is an instance of unsupervised neural algorithm and it measures the temporal separation between successive neural events. If we have one reference and one stimulus signal, then when $u_1$ precedes the $u_2$, the weight $w_2$ decreases and when the reverse happens, $w_2$ increases. This is a signature of heterosynaptic plasticity, where the evolution of synaptic strength are triggered by the activity in other pathways.

\begin{figure}
	\centering
	\includegraphics[height=8cm,width=9cm]{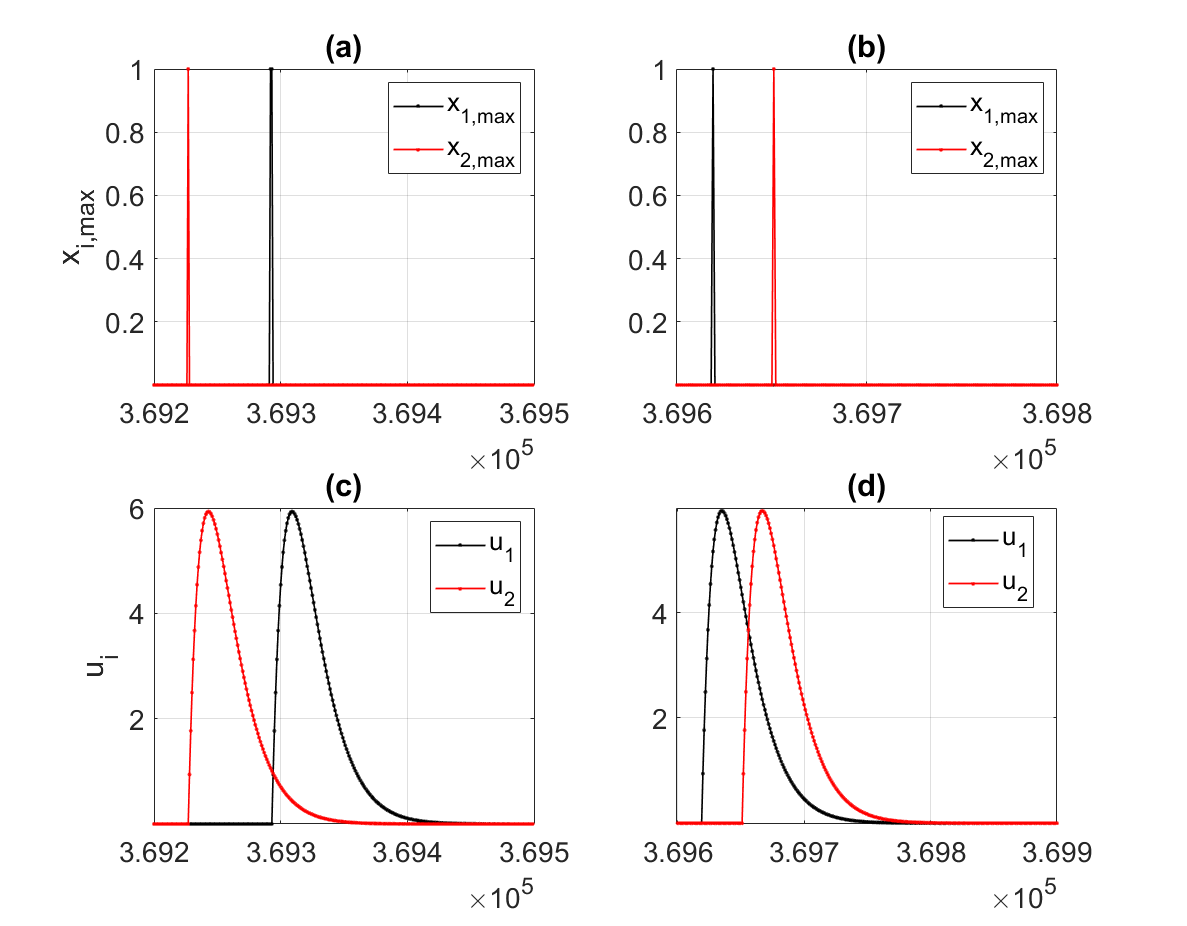}
	\caption{(a-b) Temporal maxima and (c-d) filtered temporal maxima of the time-series of the gain and loss oscillators for $A_0=0.274$.}
\end{figure}

\begin{figure}
	\centering
	\includegraphics[height=7.5cm,width=7cm]{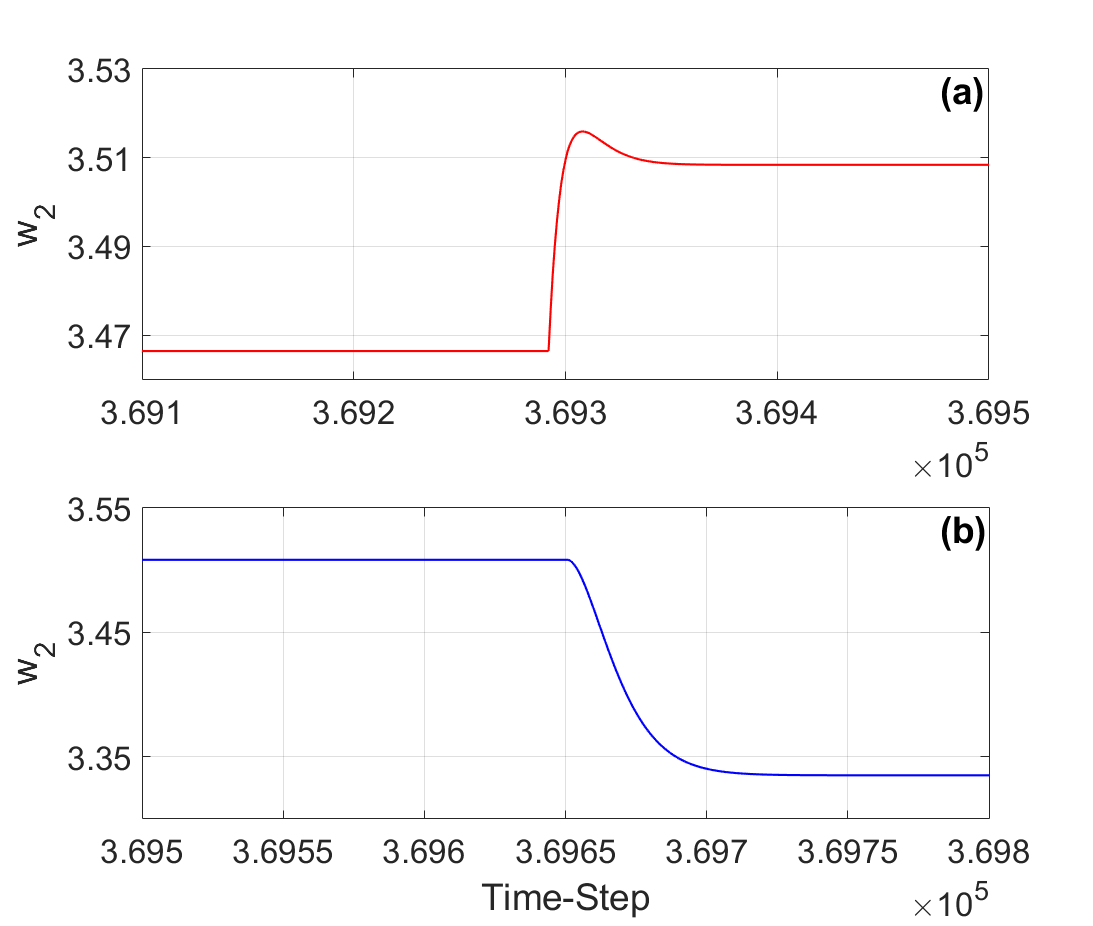}
	\caption{Temporal evolution of the weight $w_2$ in the temporal regime (a) $3.961\times10^5$ to $3.965\times10^5$ and (b) $3.965\times10^5$ to $3.968\times10^5$. Other parameters are, $Q=0.51$, $c=1$ and $f=0.01$.}
\end{figure}

\begin{figure}
	\centering
	\includegraphics[height=9cm,width=8cm]{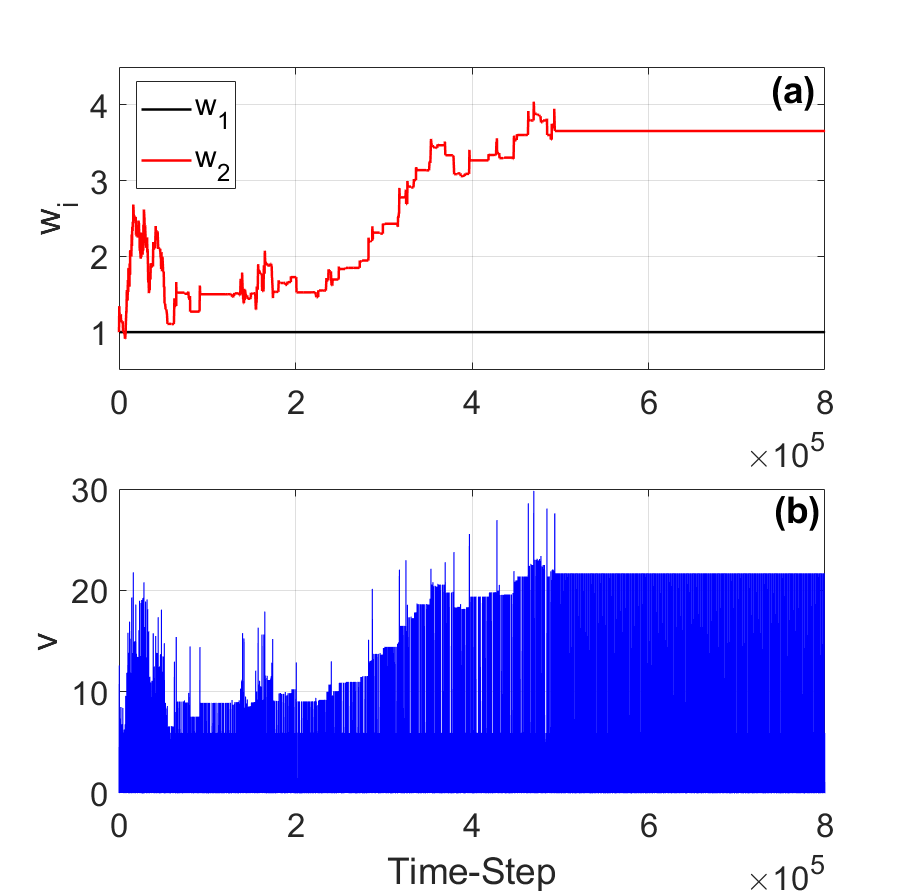}
	\caption{(a) Temporal evolution of the weights $w_1$ and $w_2$ and (b) the output $v$ for $A_0=0.274$.}
\end{figure}

In our system, we have two oscillators with continuous-time evolution. We consider the time-series of the gain oscillator $x_1$ as the reference signal and that of the loss oscillator $x_2$ as the stimulus signal. Now, the first task is to evaluate the temporal maxima in the time-series, where we used the technique employed in plotting the bifurcation diagram in Fig. 2 and Fig. 4. In addition, we shall consider the time-series of transiently chaotic dynamics of our system in discrete time-steps. If a temporal maximum was evaluated, we consider that as $1$ and if no temporal maximum was evaluated, we take that as $0$ in the time-series for temporal maxima. Afterwards, to increase the temporal overlap in the spiking signals, the raw spiking signal $x_{i,max}$ (which contains the data of the temporal maxima) is filtered and the function given below is employed.

\begin{subequations}
	\begin{align}
		& h(n) = \frac{1}{dc} e^{an} sin(dn)
	\end{align}
\end{subequations}

where $n$ is the discrete time-step, $a=-\pi f/Q$ and $d=\sqrt{(2 \pi f)^2-a^2}$. Here, $c>0$ is a parameter to tune the amplitude of the filtered signal, $f$ is frequency of the filtered normalized to unit sampling rate and $Q>0.5$ is the decay rate. Our configuration is a two oscillator system and so, we shall use two filters for the raw signals containing information of the spiking dynamics to transform them into the activation signals $u_1$ and $u_2$. The output of this learning procedure will be generated as a linear superposition of the filtered signals as follows

\begin{subequations}
	\begin{align}
		& v = w_1 u_1 + w_2 u_2
	\end{align}
\end{subequations}

Fig. 10 depicts two scenarios in the evolution of the temporal maxima of the two signals. In Fig. 10(a), we can see that the stimulus (red) precedes the reference (black) and in Fig. 10(b), the reference precedes the stimulus. The same could also be seen for the filtered signals in Figs. 10(c) and 10(d). Under both scenarios, it could be seen from Fig. 11(a) that when the stimulus $u_2 $ precedes the reference $u_1$ as shown in Fig. 10(c), $w_2$ could be seen to increasing in a spiking way and then, it decreases. But overall, $w_2$ could be seen to be increasing. But when $u_1$ precedes $u_2$ as shown in Fig. 10(d), $w_2$ is seen to be decreasing in Fig. 11(b). Furthermore, such spiking is behavior is observed at the time-step when the filtered signals overlap each other. From these facts, we can thus ascertain that our simulations provide us with the correct result. Now in Fig. 12(a), when we run our simulation till the $8\times10^5$ time-step, the temporal evolution of $w_2$ is erratic but on an average, it increases in the chaotic regime. But when the system starts exhibit periodic dynamics, $w_2$ becomes stationary. This means that the filtered signals of the reference and stimulus no longer overlap and the temporal separation between the two is constant. Likewise, the output $v$ also becomes periodic with no intermittent spiking as observed in the chaotic regime in Fig. 12(b). Note that a similar result was also observed in Fig. 9 where we have evaluated the Hilbert Transform of the time-series of the gain oscillator. All these findings clearly suggest that both the ICO Learning rule and Hilbert Transform serve a similar approach in the analysis of transiently chaotic dynamics in nonlinear systems.

\section{\label{sec5}Conclusion}
In conclusion, we have witnessed the period-doubling cascade to chaos with a brief window in which we observed intermittent chaotic behaviour in the system. The oscillators in this model also display transient chaotic dynamics in a certain parametric regimes, as validated by the maximal $\mathcal{FTLE}$. In addition, we observed that the Hilbert transform of the time-series exhibit a transformation from erratic to stationary as the time-series transform from chaotic to periodic. Also, we have applied the unsupervised learning rule called the ICO learning to the analysis of our time-series. When the system exhibits chaotic dynamics before transforming into periodic, there is an intermittent increment as well as decrement in the weight associated with the stimulus which means that the temporal separation between the signals changes erratically with time. But when the time-series transforms from chaotic to periodic, the weight associated with the stimulus becomes stationary signifying no overlap between the reference and the stimulus filtered signals. Hence, from our analysis, one can argue that this technique is an efficient measure in the analysis of the temporal evolution of any system and its associated transformation from aperiodic to periodic both in the experimental and computational sectors. Furthermore, it could be said that this learning procedure plays a role similar to the Hilbert transform in time-series analysis.

\section*{ACKNOWLEDGEMENTS}

J.P.D would like to acknowledge his group members - Prof. Claudio R. Mirasso, Prof. Ingo Fischer, Dr. Miguel C. Soriano, Dr. Apostolos Argyris and Dr. Silvia Ortín Gonzalez at the Instituto de Fisica Interdisciplinar y Sistemas Complejos (IFISC) of the Universitat de les Illes Balears (UIB), Spain for the valuable research experience in the context of ICO Learning. 

\bibliographystyle{elsarticle-num} 
\bibliography{cas-refs}
\begin{enumerate} [label={[\arabic*]}]
	\item D. O. Hebb, The organization of behavior; a neuropsychological theory, Wiley (1949).
	\item B. Porr, F. Wörgötter, Strongly Improved Stability and Faster Convergence of Temporal Sequence Learning by Using Input Correlations Only, Neural Computation 18 (6), 1380–1412 (2006).
	\item K. Möller, D. Kappel, M. Tamosiunaite, C. Tetzlaff, B. Porr, F. Wörgötter, Differential Hebbian learning with time-continuous signals for active noise reduction, PLoS ONE 17, 5 (2022).
	\item S. Löwel, W. Singer, Selection of intrinsic horizontal connections in the visual
	cortex by correlated neuronal activity, Science 255, 209–12 (1992).
	\item S. Royer, D. Paré, Conservation of total synaptic weight through balanced synaptic
	depression and potentiation, Nature 422, 518 (2003).
	\item C. H. Bailey, M. Giustetto, Y. Y. Huang, R. D. Hawkins, E. R. Kandel, Is heterosynaptic modulation essential for stabilizing hebbian plasiticity and memory, Nat. Rev. Neurosci. 1, 11 (2000).
	\item C. M. Bender and S. Boettcher, Real Spectra in Non-Hermitian Hamiltonians Having $\mathcal{PT}$ Symmetry, Phys. Rev. Lett. 80, 5243 (1998).
	
	\item C. E. R\"uter, K. G. Makris, R. El-Ganainy, D. N. Christodoulides, M. Segev, D. Kip, Observation of parity-time symmetry in optics, Nat. Phys. 6, 192-195 (2010).
	\item A. Guo, G. J. Salamo, D. Duchesne, R. Morandotti, M. Volatier-Ravat, V. Aimez, G. A. Siviloglou, and D. N. Christodoulides, Observation of $\mathcal{PT}$-Symmetry Breaking in Complex Optical Potentials, Phys. Rev. Lett. 103, 093902 (2009).
	\item J. Zhang and J. Yao, Parity-time–symmetric optoelectronic oscillator, Science Advances 4, 6 (2018).
	\item X. L\"u, Hui Jing, Jin-Yong Ma and Ying Wu, $\mathcal{PT}$-Symmetry-Breaking Chaos in Optomechanics, Phys. Rev. Lett. 114, 253601 (2015).
	\item W. Li, Y. Jiang, C. Li and H. Song, Parity-time-symmetry enhanced optomechanically-induced-transparency, Sci. Rep. 6, 31095 (2016).
	\item X. Xu, Y. Liu, C. Sun, and Y. Li, Mechanical $\mathcal{PT}$ symmetry in coupled optomechanical systems, Phys. Rev. A 92, 013852 (2015).
	\item A. K. Sarma, Modulation instability in nonlinear complex parity-time symmetric periodic structures, J. Opt. Soc. Am. B 31, 18611866 (2014).
	\item J. Schindler, A. Li, M. C. Zheng, F. M. Ellis, and T. Kottos, Experimental study of active LRC circuits with $\mathcal{PT}$ symmetries, Phys. Rev. A 84, 040101(R) (2011).
	\item M. Sarisaman, Unidirectional reflectionlessness and invisibility in the TE and TM modes of a 
	$\mathcal{PT}$-symmetric slab system, Phys. Rev. A 95, 013806 (2017).
	\item J. P. Deka and A. K. Sarma, Highly amplified light transmission in a parity-time symmetric multilayered structure, Appl. Opt. 57, 1119 (2018).
	\item Z. Lin, H. Ramezani, T. Eichelkraut, T. Kottos, H. Cao, and D. N. Christodoulides, Unidirectional Invisibility Induced by $\mathcal{PT}$-Symmetric Periodic Structures, Phys. Rev. Lett. 106, 213901 (2011).
	\item L. Feng, Y. Xu, W. S. Fegadolli, M. Lu, J. E. B. Oliveira, V. R. Almeida, Y. Chen and A. Scherer, Experimental demonstration of a unidirectional reflectionless parity-time metamaterial at optical frequencies, Nat. Mater. 12, 108-113 (2013).
	\item S. Assawaworrarit, X. Yu and S. Fan, Robust and efficient wireless power transfer using a switch-mode implementation of a nonlinear parity–time symmetric circuit, Nat. 546, 387 (2017).
	\item A. Regensburger, M. -Ali Miri, C. Bersch, J. N\"ager, G. Onishchukov, D. N. Christodoulides, and U. Peschel, Observation of Defect States in PT-Symmetric Optical Lattices, Phys. Rev. Lett. 110, 223902 (2013).
	\item A. Regensburger, C. Bersch, M. -Ali Miri, G. Onishchukov, D. N. Christodoulides and U. Peschel, Parity–time synthetic photonic lattices, Nat. 488, 167 (2012).
	\item M. -Ali Miri, A. Regensburger, U. Peschel, and D. N. Christodoulides, Optical mesh lattices with 
	$\mathcal{PT}$ symmetry, Phys. Rev. A 86, 023807 (2012).
	\item M. Wimmer, A. Regensburger, M. -Ali Miri, C. Bersch, D. N. Christodoulides and U. Peschel, Observation of optical solitons in $\mathcal{PT}$-symmetric lattices, Nat. Comm. 6, 7782 (2015).
	\item J. P. Deka and A. K. Sarma, Chaotic dynamics and optical power saturation in parity–time ($\mathcal{PT}$) symmetric double-ring resonator, Nonlinear Dyn. 96, 565 (2019).
	\item T. T\'el, The joy of transient chaos, Chaos 25, 097619 (2015).
	\item Y. -C. Lai and T. T\'el, \textit{Transient Chaos: Complex Dynamics on Finite-Time Scales} (Springer, New York, 2011).
	\item C. Grebogi, E. Ott, and J. Yorke, Chaotic Attractors in Crisis, Phys. Rev. Lett. 48, 1507 (1982).
	\item C. Grebogi, E. Ott, and J. Yorke, Crises, sudden changes in chaotic attractors, and transient chaos, Physica D 7, 181 (1983).
	\item C. Dettmann and G. Morriss, Hamiltonian reformulation and pairing of Lyapunov exponents for Nosé-Hoover dynamics, Phys. Rev. E 55, 3693 (1997).
	\item J. R. Crutchfield and K. Kaneko, Are Attractors Relevant to Turbulence?, Phys. Rev. Lett. 60, 2715 (1988).
	\item F. T. Arecchi and F. Lisi, Hopping Mechanism Generating 1/f Noise in Nonlinear Systems, Phys. Rev. Lett. 49, 94 (1982).
	\item R. Hilborn, Quantitative measurement of the parameter dependence of the onset of a crisis in a driven nonlinear oscillator, Phys. Rev. A 31, 378 (1985).
	\item R. Rollins and E. Hunt, Intermittent transient chaos at interior crises in the diode resonator, Phys. Rev. A 29, 3327 (1984).
	\item F. T. Arecchi, R. Meucci, G. Puccioni, and J. Tredicce, Experimental Evidence of Subharmonic Bifurcations, Multistability, and Turbulence in a Q-Switched Gas Laser, Phys. Rev. Lett. 49, 1217 (1982).
	\item D. Dangoisse, P. Glorieux, and D. Hennequin, Laser Chaotic Attractors in Crisis, Phys. Rev. Lett. 57, 2657 (1986).
	\item F. Papoff, D. Dangoisse, E. Poite-Hanoteau, and P. Glorieux, Chaotic transients in a $CO_2$ laser with modulated parameters: Critical slowing-down and crisis-induced intermittency, Opt. Commun. 67, 358 (1988).
	\item R. Leven, B. Pompe, C. Wilke, and B. Koch, Determining Lyapunov exponents from a time series, Physica D 16, 285 (1985).
	\item S. Sabarathinam, K. Thamilmaran, Transient chaos in a globally coupled system of nearly conservative Hamiltonian Duffing oscillators, Chaos, Solitons and Fractals 73, 129 (2015).
	\item A. Uchida, Optical Communication with Chaotic Lasers: Applications of Nonlinear Dynamics and Synchronization (John Wiley $\&$ Sons, 2012).
	\item F. Christiansen and H. H. Rugh, Computing Lyapunov spectra with continuous Gram - Schmidt orthonormalization, Nonlinearity 10, 1063 (1997).
	\item I. S. Gomez, Lyapunov exponents and poles in a non Hermitian dynamics, Chaos, Solitons and Fractals 99, 155 (2017).
	\item J. P. Deka, A. K. Sarma, A. Govindarajan, M. Kulkarni, Multifaceted nonlinear dynamics in $\mathcal{PT}$-symmetric coupled Liénard oscillators, Nonlinear Dyn 100, 1629 (2020).
	\item B. Porr, F. Wörgötter, Isotropic sequence order learning, Neural Comput 15, 831 (2003).
	\item S. Ortín, M. C. Soriano, C. Tetzlaff, F. Wörgötter, I. Fischer, C. R. Mirasso and A. Argyris, Implementation of input	correlation learning with an optoelectronic dendritic unit, Front. Phys. 11, 1112295 (2023).
\end{enumerate}

\end{document}